\documentclass[a4paper,aps,amsmath,amssymb,showpacs,10pt,twocolumn]{revtex4}
\usepackage{graphicx}
\usepackage{graphics,color}
\begin{document}
\title{Double criticality of the SK-model at $T=0$.}
\author{R. Oppermann$^{1}$, M.J. Schmidt$^{1}$, D. Sherrington$^2$}
\address{$^1$ Institut f. Theoretische Physik, Universit\"at
W\"urzburg, Am Hubland, 97074 W\"urzburg, FRG}
\address{$^3$ Rudolf Peierls Centre for Theoretical Physics, University of Oxford,
1 Keble Road, Oxford OX1 3NP, UK}
\date{march 23, 2007}
\begin{abstract}
Numerical results up to 42nd order of replica symmetry breaking (RSB) are used to
predict the singular structure of the SK spin glass at $T=0$.
We confirm predominant single parameter scaling and derive corrections for the
$T=0$ order function $q(a)$, related to a Langevin equation with pseudotime $1/a$.
$a=0$ and $a=\infty$ are shown to be {\it two critical points} for $\infty$-RSB,
associated with {\it two discrete spectra of Parisi block size ratios}, attached to a
continuous spectrum. Finite-RSB-size-scaling, associated exponents, and $T=0$-energy are
obtained with unprecedented accuracy.
\end{abstract}
\maketitle
The low temperature limit usually simplifies considerably the properties
of magnetically ordered phases.
Research in recent decades has however shown that frustrated systems
can have rich behaviour even at $T=0$.
Spin glasses \cite{MPV} are an extreme example from condensed matter,
while others are a feature of computer and information science in problems such as
hard satisfiability and error-correcting codes.
In particular, even the potentially soluble infinite-range Ising
spin glass model of Sherrington and Kirkpatrick \cite{SK} has left open many
puzzling questions.
Parisi devised an ansatz \cite{Parisi3} for the order parameter of the SK-model,
based on an infinite hierarchy of so-called replica symmetry breakings and related
hierarchically to the distribution of overlaps of metastable solutions \cite{Parisi5}.
The determining equations for this ansatz have recently been rigorously
proven to be exact \cite{Talagrand}, but its explicit solution remains elusive.
Also only recently has the $T=0$ SK problem been recognized as a critical
one-dimensional theory \cite{ro-ds-prl2005,Pankov}.

In view of the paradigmic role that the SK-model has played in the understanding
and development of the statistical physics of complex systems, together with the
potential that further comprehension of its subtleties has for extensions to other
more-complicated systems in many fields of science, especially those involving
zero- (or effectively zero-)temperature replica-symmetry-breaking
\footnote{Or for many hierarchically structured metastable phases}
transitions, it seems important to pursue the better understanding of $T=0$ RSB in
the SK model. This letter is concerned with such a study and the exposure of several
novel features, including new critical spectra, invariance points and quasi-dynamics.

Parisi's order parameter is a function $q(x,T)$ on an interval $0 \leq x \leq 1$,
the limit of a stepwise function ${q_{i}(T), x_{i}(T)}$ determined by extremization
of a free energy. It provides the hierarchical distribution of pure state overlaps
$P(q)$ through $P(q) = dx/dq$ \cite{Parisi5}.
Parisi's original work considered numerically an approximation with
a small finite number of steps, but most recent studies of the SK model have been
based on self-consistent solutions for his later non-trivial continuous order function,
typically perturbatively in the deviation from the finite-temperature phase transition.
Here the analysis is considered explicitly at $T=0$ using very accurate studies
of a very large sequence of RSB orders.

In the limit of zero temperature Parisi's order function may be
replaced by $q(a), 0 \leq a\leq \infty$,
$q(a) = \lim_{T \rightarrow 0} q_{Parisi}(aT)$ \cite{ro-hf}.
Already, in recent work to 5RSB \cite{ro-ds-prl2005} we observed a predominant
single parameter scaling for $q(a)$ and used this to predict a
simple order function $q(a)={\sqrt{\pi} a/(2\xi)}\hspace{.05cm}{\rm erf}({\xi/a})
= {_1}F_1({1/2},{3/2},-\xi^2/a^2)$
where a single correlation `length' (in RSB-space) $\xi\approx1.13$
characterized the non-trivial scaling dependence on $a$
\footnote{${_1}F_1(a,b,c)$ is the confluent hypergeometric function}.
Now, by means of redesigned numerical procedures and analytical transformations we have
been able to improve the previous $10$-dimensional extremization ($5$-RSB) up to an
$84$-dimensional extremization of the SK-energy ($42$-RSB). To the best of our knowledge,
this is by far the highest order calculation of RSB, allowing for the first time
finite `size' scaling on a rather large one-dimensional `lattice' with up to
$42$ `sites', where each site stands for one RSB-order
\footnote{The new procedure reduces the naively `exponentially hard' problem to
one where the CPU time grows polynomially with RSB-order.
Details will be given elsewhere.}.
This has enabled many hitherto unknown interesting new features to be identified.

Let us recall the SK Hamiltonian
${\cal H}=\sum_{i<j} J_{ij} \sigma_i \sigma_j$, $\hspace{.2cm} \sigma=\pm1
\label{eq:SK-model}$,
with Gaussian-distributed $J_{ij}$ independently distributed over all
pairs of sites with mean zero and variance $N^{-1}$. The $T=0$-limit \cite{ro-hf}
of the free energy can be written as the extremization with respect to the
Parisi order parameter plateaux $q_i$ and steps $a_i$ of
%
\begin{equation}
E_\kappa=\sum_{i=1}^\kappa \frac{a_i}{4}(q_i^2-q_{i+1}^2)-
\frac{\hat{{\cal T}}_{\kappa+1}^{(1)}}{a_\kappa}
\ln
(
\prod_{i=\kappa}^2 \hat{{\cal T}}_{i}^{(r_{i})}\hat{{\cal T}}_1^{(a_1)} e^{|h_1|}
)
\label{eq:T=0-energy}
\end{equation}
%
where $\hat{\cal T}_i\equiv\hat{\cal T}_{i+1,i}$ acts like a RSB-transfer matrix as
$\hat{{\cal T}}_{i+1,i}^{(r_i)} f(h_i)\equiv\frac{1}{\sqrt{2\pi\delta q_i}}\int dh_i
\exp(-\frac{1}{2\delta q_i}(h_{i+1}-h_i)^2)f(h_i)^{r_i}$ with $\delta q_i\equiv q_{i}-q_{i+1}$,
$r_{i}\equiv a_{i}/a_{i-1}$, $q_1=1, h_{\kappa+2} = 0$.

\begin{figure*}
\hspace{-.6cm}
\resizebox{.95\textwidth}{!}{%
\includegraphics{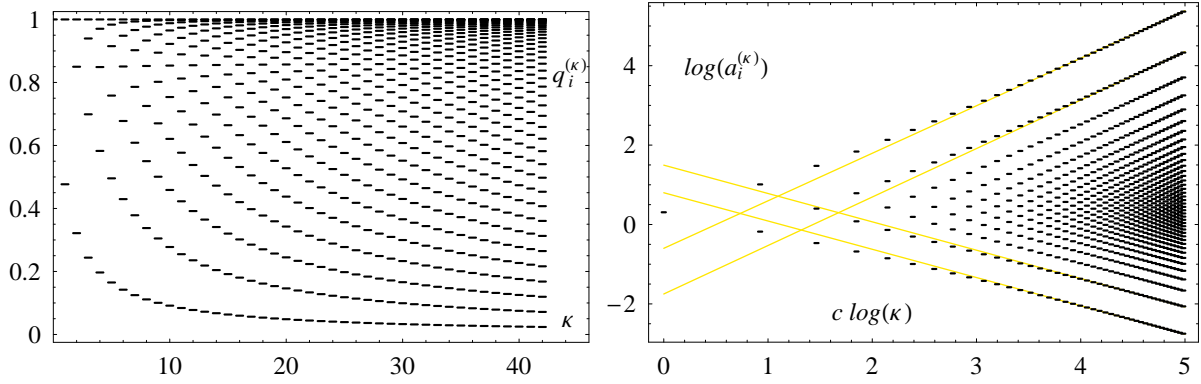}}
\caption{\label{fig:q-a-spectra}$q_i^{(\kappa)}$-spectra (left) are plotted
versus RSB-order $\kappa$; $a$-level spectra (right) are displayed by
$\log(a^{(\kappa)}_i)$ versus $c\hspace{.1cm}\log(\kappa)$ with $c\approx 4/3$;
asymptotic linear behavior of diverging $\log(a^{(\kappa)}_i)$ and a dense regime for
$a_i=O(\kappa^0)$ are observed as $\kappa\rightarrow\infty$.}
\end{figure*}

\begin{figure*}
\hspace{-.8cm}
\resizebox{0.95\textwidth}{!}{%
\includegraphics{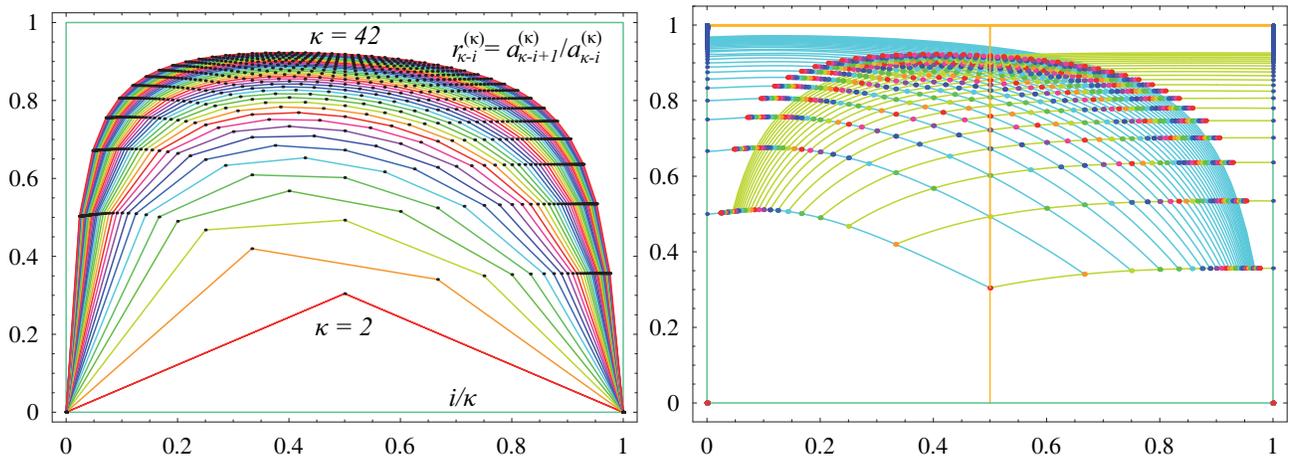}}
\caption{\label{fig:ratios}Left figure shows results for Parisi block size ratios
$r^{(\kappa)}_{\kappa-i}\equiv a^{(\kappa)}_{\kappa-i+1}/a^{(\kappa)}_{\kappa-i}$
(black dots) normalized to the unit interval $0\leq {i / \kappa} \leq 1$, connected by lines with
period-$10$ alternating colors from $2$-RSB (red, bottom) to $42$-RSB (red, top);
$a_0=\infty, a_{\kappa +1}=0$.
Right figure shows extrapolations to three different spectra at $\infty$-RSB:
a discrete spectrum on line $((0,0),(0,1))$,
a continuous distribution of $r=1$ along the line $((0,1),(1,1))$,
and a second discrete spectrum from $(1,1)$ to $(1,0)$.}
\end{figure*}

The selfconsistently obtained $q$- and $a$-level distributions are
shown in Fig.\ref{fig:q-a-spectra}. High RSB-orders $\kappa$ demonstrate
several interesting features. We remark first upon the emergence of spectral bands and
the formation of a dense region of
$a$-levels close to $a \approx  \xi $. The spectral character is further elaborated
in Fig.\ref{fig:ratios} by plotting the ratios
$r^{(\kappa)}_i\equiv a^{(\kappa)}_{i+1}/a^{(\kappa)}_{i}$
on the unit interval $0\leq i/\kappa\leq1$ for all RSB-orders $\kappa$.
\begin{figure}
\resizebox{.47\textwidth}{!}{%
\includegraphics{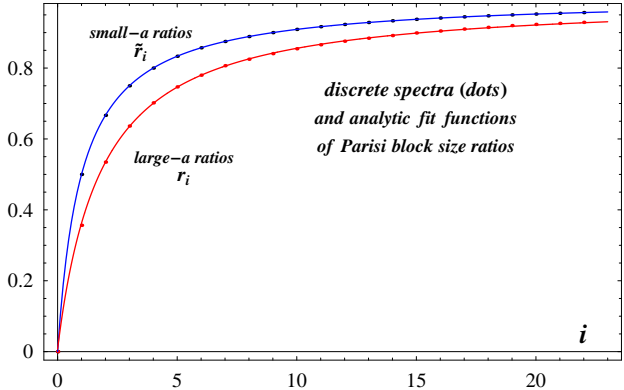}}
\caption{\label{fig:discrete-spectra}
Discrete spectra for Parisi block size ratios $r_i$ (ratios of largest $a$)
and $\tilde{r}_i\equiv r_{\kappa-(i-1)}$ (ratios of smallest $a$) as obtained
for $i=0,..,22$ in the $\kappa\rightarrow\infty$-limit (dots) compared with
fit curves interpolating (discrete) analytical predictions}
\end{figure}
Already by eye one sees that the Parisi blocksize ratios approach
characteristic limits as
$\kappa\rightarrow\infty$, further justified by fitting to Pad\'e approximants.
These demonstrate the existence of two discrete spectra on the lines
${\cal L}_1\equiv\{(0,0),(0,1)\}$ and ${\cal L}_2\equiv\{(1,0),(1,1)\}$,
respectively connected at the singular accumulation points (0,1) and (1,1)
to a continuum on $\{(0,1),(1,1)\}$.
Explicitly, the continuation to $\kappa=\infty$ yields
(i) a discrete spectrum of the form
$\tilde{r}_i\equiv lim_{\kappa\rightarrow\infty}r^{(\kappa)}_{\kappa-i}=1-1/(i+1)$
for integer-valued $i=0,1,2,...$ on ${\cal L}_1$ and
(ii) another discrete spectrum for
$r_{i}\equiv \lim_{\kappa\rightarrow\infty}r_i^{(\kappa)}$ on line ${\cal L}_2$,
well approximated by $r_i=({1+\frac65 /((i+\frac12)^{\frac25}-\frac34)^2)})^{-1/2}$
\footnote{This choice of a modified relativistic Dirac Coulomb spectrum was motivated
by several Coulomb analogies}, (see Fig 3) while
(iii) continuations at fixed finite $i/\kappa$ (see Fig.\ref{fig:ratios} right)
demonstrate the continuous spectrum (as $\kappa\rightarrow\infty$).
The two discrete spectra reflect the inequivalent non-analytic behaviors of the
order function $q(a)$ at its limiting values at $a=\infty $ and $a=0$.
\begin{figure}
\hspace{-.3cm}
\resizebox{.47\textwidth}{!}{%
\includegraphics{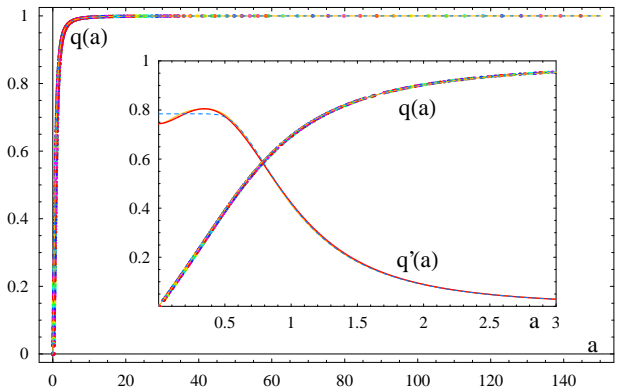}
}
\caption{\label{fig:opf} (color-online)
$10$- to $42$-RSB-data compared with analytical order function model
$q(a)={_1}F_1(\alpha,\gamma,-\xi^2/(a^2+w))a/\sqrt{a^2+w}$ for
$P_1=\{\alpha=\frac12, \gamma=\frac32, w=O(10^{-6})\}$ (dashed-blue)
and
$P_2=\{\alpha\approx0.53, \gamma\approx1.71, w\approx0.02\}$ (orange);
the insert zooms a tiny nonlinearity of $q(a)$ around $a\approx0.4$
with a maximum in $q'(a)$; the data are best reproduced for parameters $P_2$
($q'$ for $P_1$ is too flat for small $a$).
}
\label{fig:orderfunction}
\end{figure}
Fig.\ref{fig:orderfunction} shows our recent conjecture for the analytic
continuation of the $\kappa$-RSB stepped order function
$q^{(\kappa)}(a)=
\frac{\sqrt{\pi}}{2}\frac{a}{\xi_{\kappa}}{\rm erf}(\frac{\xi_{\kappa}}{\sqrt{a^2+w_{\kappa}}})$
with $\lim_{\kappa\rightarrow\infty}w_{\kappa}=0$ and $\xi_{\kappa}\rightarrow\xi\approx 1.13$
is confirmed as a good fit by the numerical solutions from $\kappa=10$ to $\kappa=42$
but on smaller $a$-scales a tiny correction is found, better resolved
by the derivative $q'(a)$ which shows a maximum near $a\approx 0.344$.
This correction can be incorporated in an improved analytical model fit function
\footnote{A minimal parameter-number choice is made which fits best all available numerical details at $T=0$;
in addition this choice confirmed the expected existence of an effective field theory with physical meaning.}
by
\begin{equation}
q^{(\kappa)}(a)=\frac{a}{\sqrt{a^2+w_{\kappa}}}
{_1}F_1(\alpha_{\kappa},\gamma_{\kappa},-\xi_{\kappa}^2/(a^2+w_{\kappa})).
\label{eq:q(a)}
\end{equation}
The flow of the variational parameters as $\kappa\rightarrow\infty$ remains
close to our previous proposal
\cite{ro-ds-prl2005}:
the limits slightly modified to $\alpha\equiv\alpha_{\infty}= 0.53\pm 0.01$,
$\gamma\equiv\gamma_{\infty}=1.71\pm0.02$, $\xi\equiv\xi_{\infty}=1.16\pm0.01$,
and $w$ monotonically falling towards zero
\footnote{$w=w_{\infty}=0$ implies $q'(a)\rightarrow0$ in an
exponentially small $a$-regime which
cannot be resolved by the numerical data so that a small but finite
$w_{\infty}$ cannot be ruled out}.

We observe that the large $a$-expansion
$q^{\infty}(a)=1-\frac{\alpha}{\gamma}\frac{\xi^2}{a^2}
+\frac{\alpha(\alpha+1)}{\gamma(\gamma+1)}\frac{\xi^4}{2a^4}+O(a^{-6})\approx
1-0.42 a^{-2}+0.16 a^{-4}+O(a^{-6})$ of Eq.\ref{eq:q(a)}
agrees well with Pankov's $q(a)=1-0.41 a^{-2}$ obtained from the
standard $\infty$-RSB differential equation
and $T\rightarrow0$ scaling \cite{Pankov}.
The deviation of $\alpha$ and $\gamma$ from the rational values $\frac12$ and $\frac32$
of our original proposal \cite{ro-ds-prl2005} leads to a small correction to a
linear rise of $q(a)$ at small $a$  and a pronounced maximum of $q'(a)$ at $a=0.344$.
This feature confirms low $T$ results of Crisanti and Rizzo \cite{CR}, obtained
perturbatively.
It is also interesting to compare our results for $q(a)$ with the predictions of
the Vannimenus-Toulouse-Parisi scaling ansatz $q_{Parisi}(x,T) = f(x/T)\leq q_{max}(T)$
\cite{PaT,VTP}, combined with the Parisi-Toulouse hypothesis \cite{PaT} and
the knowledge of the behaviour on the de Almeida-Thouless line.
The comparison is qualitatively quite good but not perfect, most noticeably
deviating at low $a$, the VTP prediction being slightly lower and more curved.

Our analysis also gives strong hints for the existence of
{\it invariance-points} in $a$-space,
i.e. points where $q^{(\kappa)}(a)$ or one of its derivatives
$\partial^{n}_{a}q^{(\kappa)}(a)$
does not vary under a change of RSB-order. They appear to be fixed
points under the
renormalization group which decimates the RSB-order \cite{ro-ds-prl2005}.
For example $q^{(\kappa)}(a)$ does not change with $\kappa$ at $a={\bar a}\approx 0.401$;
for all calculated orders the changes $q^{(\kappa+1)}(a)-q^{(\kappa)}(a)$ are negative
for $a<{\bar a}$ and positive for $a>{\bar a}$; they decay towards zero in $\infty$-RSB
without change of sign, such that the contributions accumulate.
As yet we have no physical explanation for these new critical ``a-values".
\begin{figure*}
\hspace{-.2cm}
\resizebox{.94\textwidth}{!}{%
\includegraphics{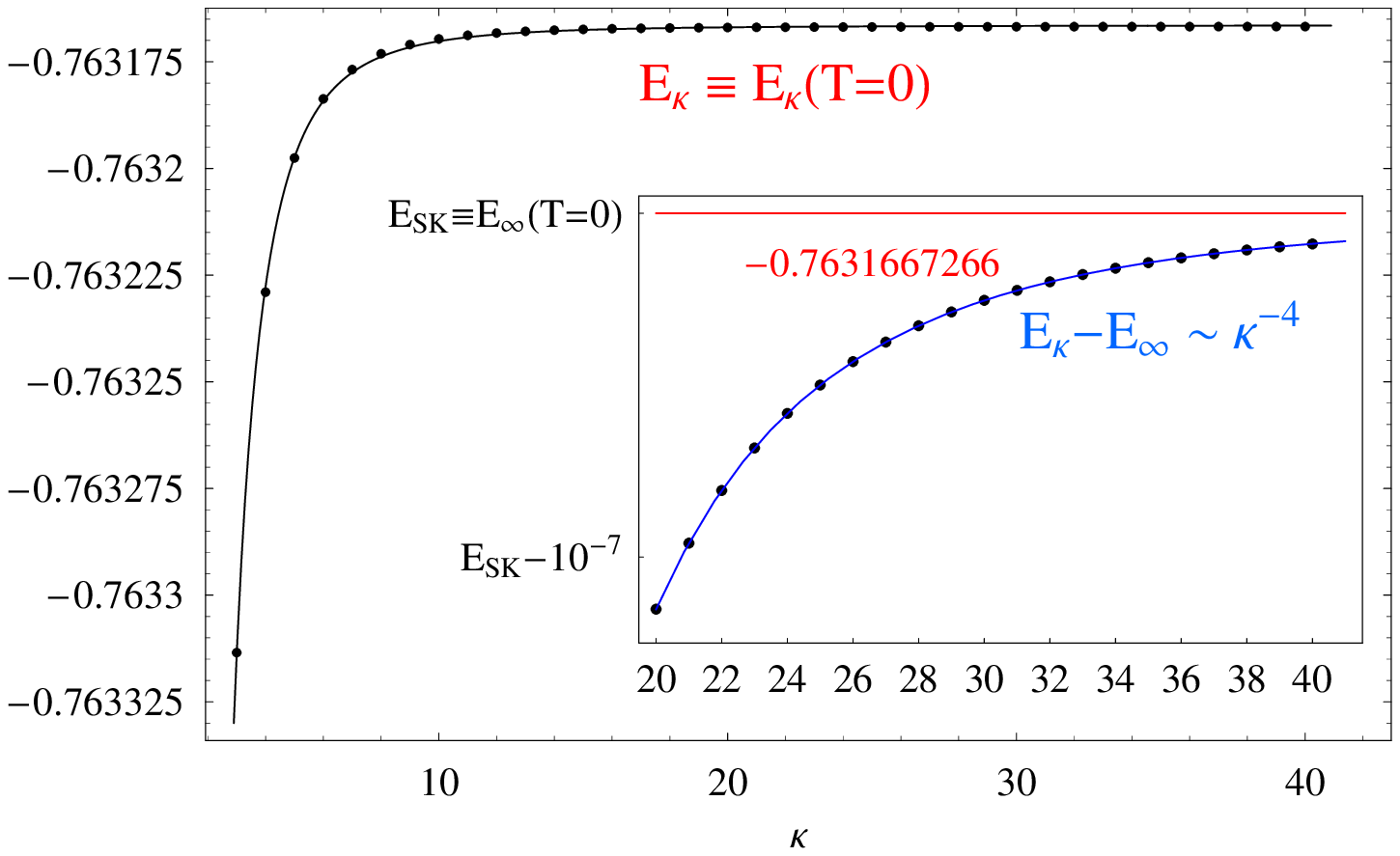}
\includegraphics{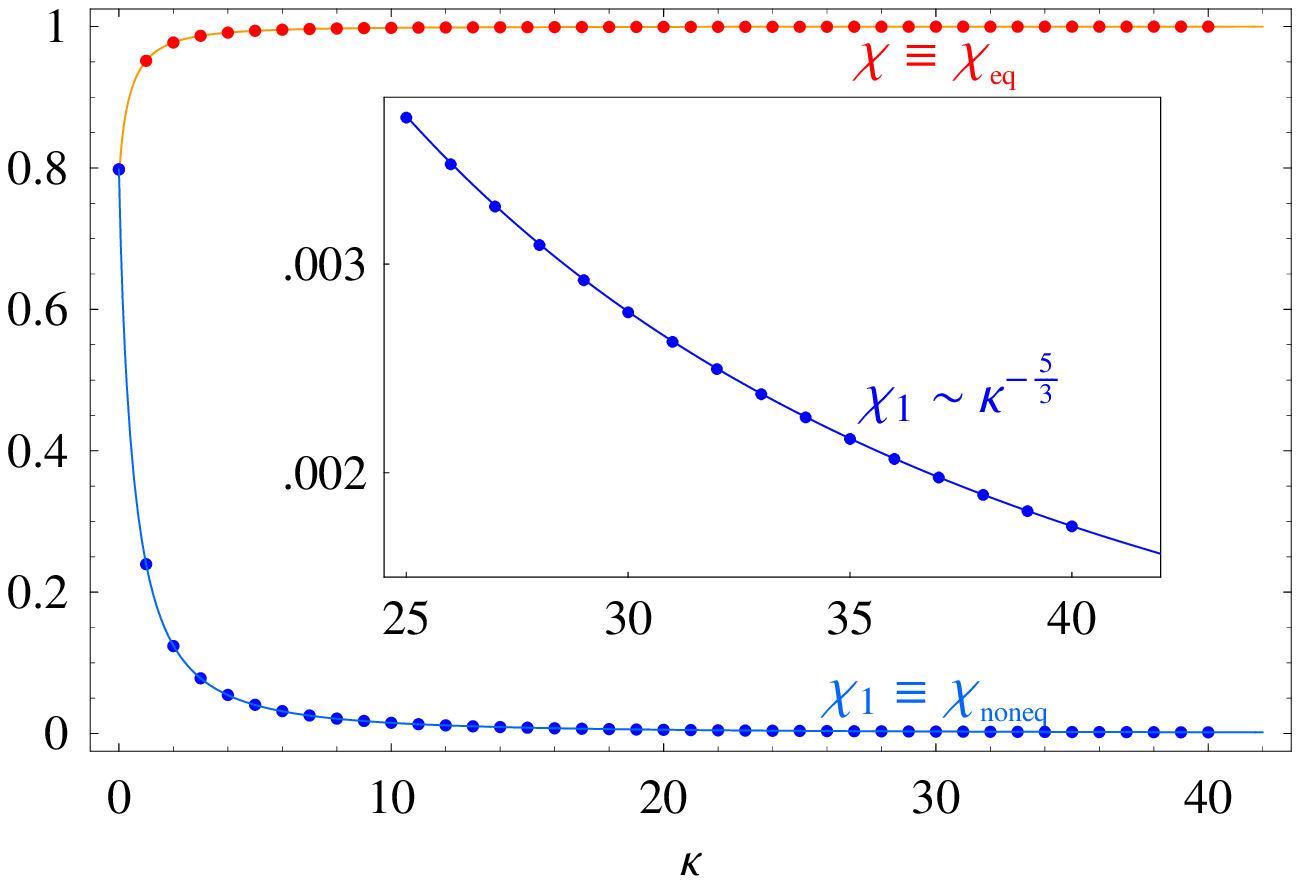}}
\caption{Flow of $T=0$ SK-energy (left) and of thermodynamic and single-state
susceptibilities $\chi^{(\kappa)}$, $\chi_1^{(\kappa)}$ (right) as a function of RSB-order
$\kappa$ towards their $\kappa=\infty$-limits $E_{SK}$ and $\chi_1=0, \chi=1$.
Inserts show pure power law fits of data in high $\kappa$-regime.}
\label{fig:E-chi-flows}
\end{figure*}

Our accurate results, $O(10^{-12})$, for such a large number of
RSB orders also enables us to
determine finite-RSB-size scaling exponents and the ground state energy
to unprecedented accuracy.
The energy of the SK-model is plotted versus RSB-order $\kappa$
in Fig.\ref{fig:E-chi-flows}.
Within the orders studied both the energy and the susceptibilities as a function of
$\kappa$ have reached the asymptotic large-$\kappa$ scaling regime,
fitting accurately the power law behaviors
\begin{equation}
1-E_{\kappa}/E_{\infty}\sim {\kappa}^{-z_E}, \hspace{.5cm}\chi_1\sim {\kappa}^{-z_{\chi}}
\end{equation}
with  $E_{\infty}=-0.7631667265(6)$, $z_{E}\approx 4$ and $z_{\chi}\approx \frac53$
and $(\chi - 1)$ vanishing exponentially
\footnote{The susceptibility $\chi_1$ derives from the variation of the SK energy
$(1)$ with respect to $q_1$ (largest-$a$ regime)}.
Fig.\ref{fig:E-chi-flows} also shows few-parameter analytic functions that
yield good fits for the full range of RSB-orders.
The results for $E_{\infty}$ and $1=\chi(T=0)=\int_0^{\infty} da\hspace{.1cm}a\hspace{.1cm}q'(a)$
are in accord with Eq.(2).

In an earlier $5$-RSB study of the $T=0$ distribution of local fields
$P(h)=N^{-1}\sum_i\langle{\delta(h-\sum_j{J_{ij}}\sigma_j)}\rangle$ there appeared to
be a deviation from linearity in $|h|$ at very small $h$ \cite {ro-ds-prl2005}.
In the light of the extension to $40$-RSB, we are able to show
that this is an artefact of finite RSB order and the power law
$P(h)\approx0.3|h|$ for small $h$ \cite{TAP, Thomson} is confirmed.
In any specific finite-RSB order, small oscillations are observed around
$P'(h)\approx0.3$ near $h=\chi_1$ but they shrink to zero as
$\chi_1\rightarrow0$ in the $\infty$-RSB limit.
Thus in the present context the pseudogap behavior $P(h)\sim|h|$ can be viewed as being generated
by the decay of finite-$\kappa$ RSB gaps as $\kappa\rightarrow\infty$.

One may further relate our results to a pseudo-dynamical field theory. The
order function $q(a)$ in Eq.(2) can be recognised for $w=0$ as the solution
of the confluent hypergeometric differential equation
$(a^4/\xi^2)\partial_a^2 q(a)+((3-2\gamma)a^2/\xi^2-2)a\partial_a q(a)+4\alpha q(a)=0$,
or given by
\begin{eqnarray}
& &\hspace{-.8cm}\frac{\partial}{\partial(\xi/a)}\phi(a)=\frac{\delta{\cal H}}{\delta\phi(a)},
\hspace{.2cm}
\phi(a)=-\frac{1}{\sqrt{2\alpha}}\frac{\partial}{\partial(\xi/a)}\hspace{.1cm}\log(q(a))\nonumber\\
& &\hspace{-.8cm}{\cal H}=\sqrt{8\alpha}\hspace{.05cm}\sinh(\phi(a))
-\left((\gamma-1/2)a/\xi+\xi/a\right)\phi^2(a)
\label{SK_eom}
\end{eqnarray}
provided $\sinh(\phi)$ is truncated at third order. In fact, however,
the numerical solution of the $\sinh(\phi)$-model (\ref{SK_eom})
provides as good a fit to our data as Eq.($\ref{eq:q(a)}$)
and thus mimics similarly well the SK-model at $T=0$.
The relationship to dynamics follows from an inverse relationship
between the Parisi `length' $x$
(or $a$ as here) and the timescale to explore the corresponding clusters of pure
states in dynamics, with the characteristic timescales increasing sharply as $x$
is reduced \cite{Sompolinsky}. Taking $1/a$ as a pseudo-time Eq.(\ref{SK_eom}) is
identified as a (Langevin) pseudo-dynamical equation.
The parameters $\alpha$ and $\gamma$, which model
the order function $q(a)$ analytically in between the singular points $a=0$, $a=\infty$,
appear as two new coupling constants in the Hamiltonian.
We also recall here the known relationship between replica-symmetry breaking and slow dynamics,
much in analogy with the Halperin-Hohenberg theory \cite{HHT}
of slow-dynamical (and spatially-critical)
behavior.

In summary, we have obtained several new results for the
paradigmic Sherrington-Kirkpatrick spin-glass at
$T=0$:
(i)
we have found that the Parisi order function can be represented by a
1D field theory with two inequivalent critical limits $a=0$ and $a=\infty$,
the order parameter $q(a)$ showing critical behavior in both limits
of long ($a=0$) and short ($a=\infty$) pseudotimes.
The key issue for criticality is the presence of two
discrete spectra for the Parisi block size ratios
(at $a=0$ and at $a=\infty$), attached to a continuous spectrum for finite $a$.
(ii)
All orders of RSB up to 42nd order have been solved to great numerical accuracy
to enter deep into the large-RSB scaling regime, enabling us to obtain new power laws,
the SK-energy up to $O(10^{-10})$, and corrections to single parameter scaling in a
hypergeometric order function model.

It would be interesting to extend to systems, both range-free and finite-range,
and to include real temporal dynamics and quantum effects. As examples of
potential non-trivial range-free
extensions one might consider K-SAT or LDPC code problems under dynamical algorithmic
optimization using either classical or quantum dynamical
annealing
\footnote{In particular, although several effectively mean-field (range-free)
problems, with higher-order ($p>2$) interactions have 1RSB thermodynamics
near their onsets from their finite-temperature ergodic regimes, they normally
exhibit full RSB at $T=0$, while finite connectivity also increases
the size of the order parameter space.}.
The existence of true RSB in finite-range problems at equilibrium is
still controversial but our study
could potentially usefully extend current Ginzburg-Landau-Wilson
functional integral theory\cite{DG}.

This work was supported by DFG through grant Op28/5-2 and SFB410-D5,
by EPSRC through grant GR/R83712/01, and by ESF through program SPHINX.
We thank Ch. Gould and L. Molenkamp for support and
M. M\"uller and A. Sportiello for useful discussions.

\end{document}